\documentclass[aps,prx,showpacs,twocolumn,groupedaddress,superscriptaddress]{revtex4-1}

\usepackage{times,xspace}
\usepackage{graphicx,graphics,color,epsfig}
\usepackage{amsmath}
\usepackage{amssymb}
\usepackage{amsfonts}
\usepackage{amsfonts}
\usepackage{epstopdf}
\usepackage{bm}
\usepackage{hyperref}
\usepackage{color}
\usepackage{float}
\usepackage[normalem]{ulem}
\usepackage{comment}
\usepackage{soul}

\def\be{\begin{equation}}
\def\ee{\end{equation}}
\def\ba{\begin{eqnarray}}
\def\ea{\end{eqnarray}}

\newcommand{\figref}[1]{Fig.~\ref{#1}}

\begin{document}

\title{Photo-induced charge, spin, and orbital order in the two-orbital extended Hubbard model}

\author{Sujay Ray}
\author{Philipp Werner}
\affiliation{Department of Physics, University of Fribourg, 1700 Fribourg, Switzerland}

\date{\today}
 
\begin{abstract}
Nonequilibrium control of electronically ordered hidden phases may lead to the development of ultrafast switches and memory devices. In this study, we demonstrate tunable hidden orders in the photo-doped two-orbital extended Hubbard model. Using steady-state nonequilibrium dynamical mean field theory, we clarify the coexistence and interplay of nonthermal charge, spin, and orbital order. The hidden state at low effective temperature and sufficiently high photo-doping is reminiscent of Kugel-Khomskii order in the two-orbital Hubbard model at $\frac{1}{4}$ and $\frac{3}{4}$ filling, but it emerges out of a nonequilibrium charge ordered state and exhibits a different magnetic structure. A low-energy effective Hamiltonian is used to analyze the exchange processes which stabilize the nonthermal order.  
\end{abstract}
\vspace{0.5in}

\maketitle
 
\section{Introduction} The nonequilibrium manipulation of electronic orders is an active research field, driven by curiosity and potential technological applications such as ultrafast switches and storage devices \cite{ultrafast_qm,ultrafast_os,ultrafast_op,ultrafast_hid,nonth_Kugel_Khomskii}. A particularly interesting prospect is the controlled access to hidden orders, which are metastable states with properties that are distinct from the equilibrium phases of the system \cite{ultrafast_qm,ultrafast_op,lat_struc,Cavalleri2018,ph_noneq,opt_exint}. Correspondingly, over the last decade, there have been numerous experimental and theoretical efforts to describe and understand nonequilibrium properties of correlated electron systems \cite{VO2_dynamics,ferro_chrgtrnsfr_dynamics,mag_switch,Qint_Mott,opt_magcont,1d_photodoping, noneq_AFM,photo_Mott,photo_Hund_exciton,nonthermal_Mott,neq_GW}. Among the studied laser-excited strongly correlated materials, Mott insulating systems are particularly promising candidates \cite{ph_noneq}, since the Mott gap prevents a rapid heating and thermalization of the system, which enables the emergence of a plethora of hidden orders ranging from superconducting states \cite{yuta_etasc1,yuta_etasc2,eta_jiajun,tri_chiral,kaneko1,kaneko2,eta_spintrip,dex_SC_ferro} to magnetic orders \cite{Ishihara_DE1,Ishihara_DE2,dex_SC_ferro}, charge orders \cite{yuta_etasc1,yuta_etasc2}, Kugel-Khomskii spin and orbital orders \cite{nonth_Kugel_Khomskii},  as well as excitonic orders \cite{exciton_insulator}. These hidden orders can be accessed by photo-doping the Mott insulators with ultra-short (femtosecond) laser pulses and they can form on sub-picosecond timescales. Hence, in large-gap Mott insulators with slow recombination of the photo-carriers \cite{Sensarma2010,Eckstein2011,Strohmaier2010,Lenarcic2014}, these hidden orders are metastable and can be described within a non-equilibrium steady state formalism \cite{NESS_martin,Mrtin_inchworm}. Since the photo-induced hidden orders are intricately linked to the presence of photo-carriers, they are distinct from thermal phases, and they can be manipulated with ultra-short laser pulses. 

In this work, we explore the possibility of emergent spin and orbital order in the photo-doped extended two-orbital Hubbard model ($U$-$V$ model). Nonequilibrium phases of the extended Hubbard model have previously been investigated for the single-band case, where a photo-induced bi-excitonic order has been found in the presence of next-nearest neighbor interactions  \cite{biexciton_exhubbard}, while other studies revealed a photo-induced charge ordered state above a critical nearest-neighbor interaction $V_{\text{c}}$ \cite{yuta_etasc1,yuta_etasc2}. In general, this charge ordered state competes with staggered superconducting order, which appears below $V_\text{c}$. In the two-orbital model, additional orders such as spin and orbital orders play a crucial role, in contrast to the single-orbital model with spinless charge excitations (doublons and holons). Our study shows that the spin and orbital orders in the photo-doped half-filled two-orbital Mott insulator do not compete with the charge order. Rather, they complement each other, which leads to the intriguing possibility of a photo-induced coexistence of spin, orbital and charge order. 

An example of an intertwined spin and orbital order is the Kugel-Khomskii order in the $1/4$ or $3/4$ filled two-orbital Hubbard model \cite{Kugel_Khomskii,Kugel_Khomskii_1}. In the photo-doped model, the Kugel-Khomskii order has been shown to melt after the application of an ultra-short laser pulse \cite{nonth_Kugel_Khomskii}. While the melting of magnetic and orbital order is expected in non-equilibrium phases because of the disordering of antiferro-type orders by mobile charge carriers \cite{ph_noneq}, as well as heating, the emergence of these orders is rare in non-equilibrium phases. The latter is however crucial for the development of ultrafast electronic devices.

\begin{figure}[b]
\includegraphics[width=0.48\textwidth]{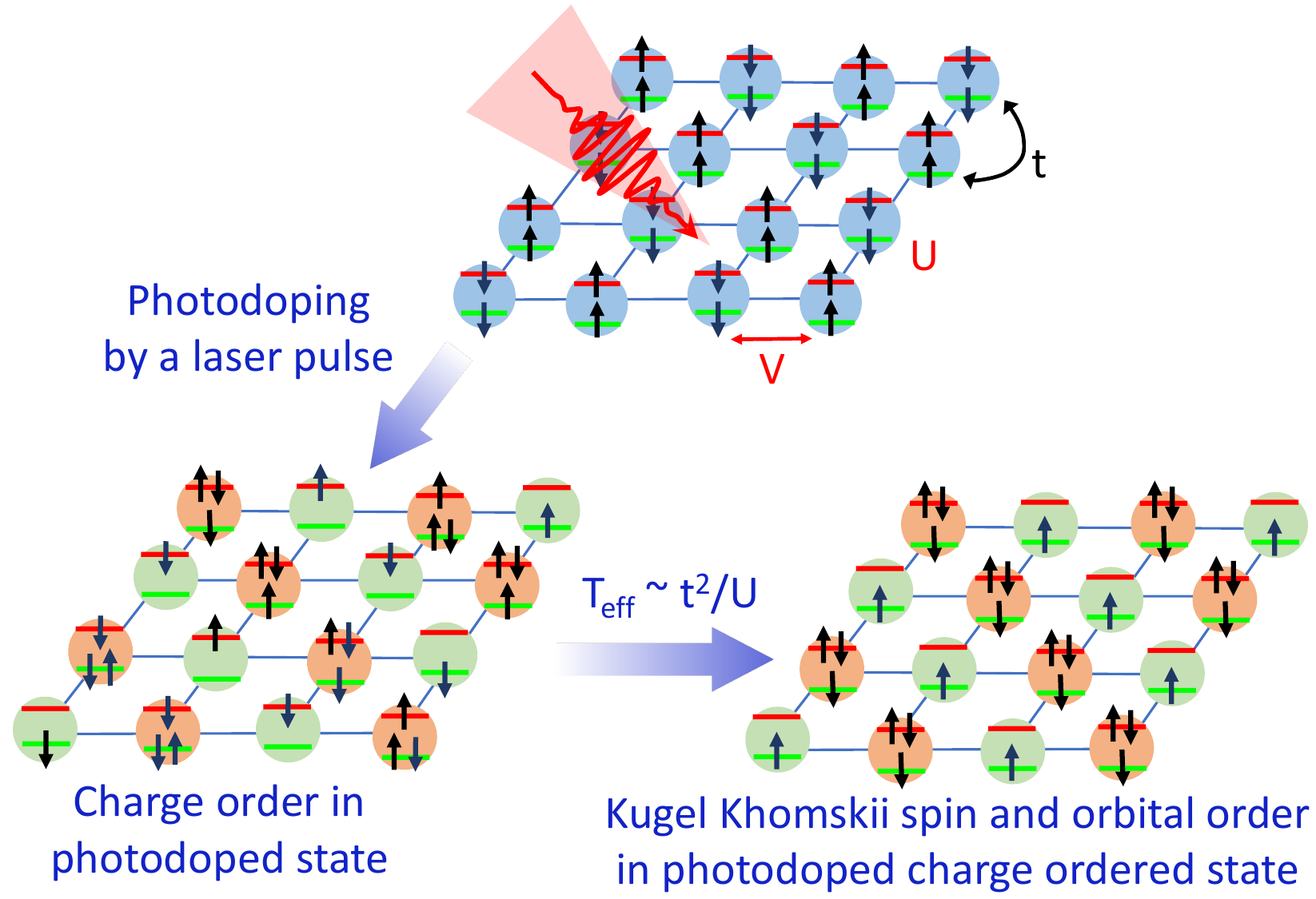}
\caption{ Schematic illustration of the photo-excited ordered states in a two-orbital extended Hubbard model. The charge carriers created by the laser pulse become charge ordered because of the nearest neighbor repulsive interaction $V$. If the effective temperature $T_\text{eff}$ of the photo-doped system is lowered below $\sim t_{h}^{2}/U$, spin and orbital order appears in addition to the charge order.
}
\label{fig1}
\end{figure}

Recently, we have demonstrated photo-induced ferromagnetic order in a two-orbital Hubbard model with orbital-dependent hopping amplitudes \cite{dex_SC_ferro}. But emergent orbital order has not yet been achieved in photo-doped systems. Here, with the help of nonequilibrium DMFT calculations, we demonstrate that the charge excitations in the photo-doped half-filled two-orbital system on a bipartite lattice result in a charge ordered state as shown in ~\figref{fig1}. If the photo-doping concentration is large enough (high density of charge ordered triplons and singlons),  Kugel-Khomskii spin and orbital order can be realized in an effectively cold state. This state with multiple intertwined electronic orders is an interesting prototype of a photo-induced hidden phase. While the realization in solids may be challenging due to heating effects, cold atom systems may in the near future become a versatile platform for the study of such nonequilibrium multi-orbital physics \cite{richaud2022}. 

The rest of the paper is organized as follows: in Sec.~\ref{sec_model}, we describe our model and the non-equilibrium steady-state DMFT method. In Sec.~\ref{sec_eff_lowen_model}, we analyze the emerging Kugel-Khomskii order in our model with the help of an effective low-energy Hamiltonian. In Sec.~\ref{sec_results}, we present our equilibrium and nonequilibrium DMFT calculations, while Sec.~\ref{sec_conclusion} is a brief summary and conclusion of our findings.

\section{Model and Method}
\label{sec_model}
We consider the the two-orbital extended Hubbard model with Hund coupling and nearest-neighbor repulsive interactions,
\begin{align}
&H = -t_{\text{h}} \sum_{\left<ij\right>,\sigma} \sum_{\alpha=1,2} c^{\dagger}_{i,\alpha\sigma}c_{j,\alpha\sigma} + U\sum_{i}\sum_{\alpha=1,2}n_{i,\alpha\uparrow}n_{i,\alpha\downarrow} \nonumber \\
&\hspace{0mm} - \mu \sum_{i}\sum_{\alpha=1,2}\left(n_{i,\alpha\uparrow} + n_{i,\alpha\downarrow} \right) + (U-2J)\sum_{i,\sigma}n_{i,1\sigma}n_{i,2\Bar{\sigma}} \nonumber \\
&\hspace{0mm} + (U-3J)\sum_{i,\sigma}n_{i,1\sigma}n_{i,2\sigma} + V\sum_{\left<ij\right>}n_in_j,\label{eq_1}
\end{align}
where $c^{\dagger}_{i,\alpha\sigma}$ $(c_{i,\alpha\sigma} \nonumber)$ is the creation (annihilation) operator at site $i$, $\alpha$ and $\sigma$ denote the orbital and spin indices, $t_{\text{h}}$ is the nearest-neighbor hopping amplitude between sites $i$ and $j$, $U$ is the intra-orbital Hubbard repulsion, $J$ the Hund coupling, $V$ the nearest-neighbor Hubbard repulsion and $\mu$ the chemical potential. For $V=0$, the ground state at half-filling is a antiferro-magnetic (antiferro-orbital) state for $J>0$ ($J<0$), which is described by a SU(2) spin (orbital) order. At $1/4$ or $3/4$ filling, the gorund state corresponds to a SU(4) spin and orbital order (Kugel-Khomskii order) \cite{Kugel_Khomskii}. 

In this study, we investigate the evolution of SU(2) order to SU(4) order, driven by 
a staggered potential or photo-doping. 
Our calculations for an infinitely-coordinated Bethe lattice are based on dynamical mean field theory (DMFT) \cite{Georges_1996, Aoki_2014} with a non-crossing approximation (NCA) \cite{Keiter_1970, Eckstein_2010} impurity solver. This method should give solutions representative of high-dimensional bipartite lattices and reveal the relevant electronic orders in the large-$U$ limit. 
We first consider the effect of site-dependent chemical potential shifts by introducing two impurity models for sublattices $A$ and $B$, and applying a positive (negative) chemical potential shift to sublattice $A$ ($B$). This allows us to dope one sublattice with holes and the other sublattice with electrons, and hence to artificially create a charge-ordered state. 

While sublattice-dependent chemical potential shifts seem artificial, the resulting states have a close resemblance to the half-filled photo-doped extended Hubbard model with nonzero $V$. To simulate the photodoped systems, we employ the nonequilibrium steady state (NESS) DMFT formalism \cite{NESS_martin}. In the NESS simulation an effective photoexcited state is created by weakly coupling cold Fermion baths to the half-filled Mott system at each site. These  Fermion baths are aligned with the lower and the upper Hubbard bands, which allows to  effectively induce charge carriers (triply and singly occupied sites) into the system. Within the steady-state formalism, we can efficiently tune the density of photo-carriers by choosing the chemical potentials of the Fermion baths $\mu_b$. By changing the temperature of the Fermion baths $T_b$, it is furthermore possible to control the effective temperature of the photo-doped system, which can be defined by fitting a Fermi distribution function to the ratio of the occupied density of states and total density of states in the energy range of the Hubbard bands \cite{eta_jiajun, eta_spintrip}. This setup allows us to search for hidden orders in the photo-excited system by varying several tuning parameters like Hubbard repulsion, Hund coupling, photo-doping concentration and effective temperature.



The nearest-neighbor interaction in Eq.~\eqref{eq_1} cannot be captured by the  DMFT formalism. We treat this interaction at the mean field level and include its effect in the self-consistent solution of two coupled impurity models. In an infinitely coordinated lattice, the energies associated with the non-local terms in the Hamiltonian diverge, unless we rescale the parameters. To get meaningful solutions in the infinite-coordination limit, the hopping parameter has to be rescaled as $t_{\text{h}}^*\propto t_{\text{h}}/\sqrt{Z_n}$ \cite{Metzner1989}, and the nearest-neighbor interaction term as $V^*= V/Z_n$, with $Z_n$ the coordination number. In the following DMFT analysis, we thus define an effective hopping parameter $t_0$ related to the bandwidth.  

In a mean field picture, and for a bipartite lattice, we only need to consider the two average densities $\langle n_{A}\rangle$ and $\langle n_{B}\rangle$ for the sites on sublattice $A$ and $B$. In the impurity model for sublattice $A$, the nonlocal interaction then adds a term $H^{A}_{V} = V^* Z_n n_A\left<n_B\right> = V n_A\left<n_B\right>$, and analogously for sublattice $B$ we have $H^{B}_{V} = V n_B\left<n_A\right>$. The DMFT self-consistency equations take the form
\begin{eqnarray}
    \Delta^{A}(t,t^{\prime})=t_{0}^{2} G^{B}(t,t^{\prime}) + \sum_{b} D_{b}(t,t^{\prime}), \\
    \Delta^{B}(t,t^{\prime})=t_{0}^{2} G^{A}(t,t^{\prime}) + \sum_{b} D_{b}(t,t^{\prime}).
\label{eq_2}
\end{eqnarray}
Here $\Delta^{A(B)}$ is the hybridization function, $G^{A(B)}(t,t^{\prime})$ is the Green's function for sublattice $A$ $(B)$, and $t_{0}=W/2$, where $W$ is the half-bandwidth. $D_{b}(t,t^{\prime})$ is the contribution to the hybridization function from the Fermionic baths, which is defined in frequency space by $D_b(\omega)=$ $g^2\rho_{b}(\omega)=\Gamma \sqrt{W_{b}^{2}-(\omega - \omega_{b})^{2}}$. Here, $\Gamma=g^2/W_{b}^{2}$ is a dimensionless coupling constant, $\omega_b$ indicates the center of the energy spectrum and $W_b$ indicates the half-bandwidth of the bath $b$. Because of the steady state assumption, the hybridization functions and Green's functions depend only on the time difference $t-t^{\prime}$ and can be transformed back and forth between time and frequency space using fast Fourier transformation. 

In all our calculations we use four fermion baths with $\omega_{b}=\pm U/2, \pm 3U/2$,  $W_b=4.0$ and $\Gamma=0.044$ for efficient photo-doping and we set the local repulsion to $U=20$ and the nearest-neighbor repulsion to $V=0.3$. We measure the observables in both sublattices $A$ and $B$. The magnetic orders are defined via the local magnetization on sublattice $A/B$ as $m_{A/B}= \sum_{\alpha}(n_{A/B,\alpha\uparrow} - n_{A/B,\alpha\downarrow})$. Similarly, the orbital order is defined as $o_{A/B}=\sum_{\sigma}(n_{A/B,1\sigma} - n_{A/B,2\sigma})$. We also define charge order as $c_{A/B}=n_{A/B}-2$. The triplon density, defined as $d=\left< n_{1\uparrow}n_{1\downarrow}n_{2}+n_{1}n_{2\uparrow}n_{2\downarrow}-4n_{1\uparrow}n_{1\downarrow}n_{2\uparrow}n_{2\downarrow} \right>$, serves as a measure for the amount of photo-doping ($n_{\alpha}=n_{\alpha\uparrow} + n_{\alpha\downarrow}$ is the occupation of orbital $\alpha$).

In order to detect different orders we apply small (staggered) seed fields which couple to the order parameters of interest. In a symmetry-broken state the order parameter grows to a high value, which is almost independent of the value of the seed field. In a charge ordered phase $c_A=-c_B \neq 0$, while in an antiferro-magnetic (antiferro-orbital) phase $m_A=-m_B\neq 0$ ($o_A=-o_B\neq 0$).

\section{Effective low-energy model}
\label{sec_eff_lowen_model}
We can gain insights into the properties of the photo-doped Hubbard model by considering an effective low-energy Hamiltonian. By applying this effective  Hamiltonian to the relevant nonequilibrium multiplets (doublons, holons in the single-orbital case and triplons, singlons in the half-filled two-orbital case), we can qualitatively describe the photo-doped system. This works well in the limiting cases, e.~g., the complete photodoping limit. Such an analysis has correctly predicted $\eta$-superconducting phases and magnetic phases in photo-doped strongly-correlated systems \cite{yuta_etasc1,yuta_etasc2,eta_jiajun,eta_spintrip,dex_SC_ferro}. 

Let us start with the single-orbital extended Hubbard model, for which the effective low-energy Hamiltonian is given by \cite{eta_jiajun,eta_spintrip,SWT}
\begin{equation}
H_\text{eff} = H_\text{hop}(P_{in}P_{jn-1}) + H_{s}(P_{in}P_{jn^{\prime}})
+ H_{\eta}(P_{in}P_{jn-2}), \label{eq_3}
\end{equation}
where $P_{in}$ is the projection operator at site $i$ on states with particle number $n$ (for a single orbital $n=0,1,2$). In Eq.~\eqref{eq_3}, $H_\text{hop}=-t_\text{h}\sum_{\left<ij\right>}\sum_{\alpha} \big(c^{\dagger}_{i,\sigma} c_{j,\sigma} +h.c. \big)$, which represents the kinetic energy of the doublons and holons. This term acts on neighboring sites which differ in particle number by one. The second term is the Heisenberg spin term  $H_s=4t_{\text{h}}^2/U \sum_{\left<ij\right>} \boldsymbol{s_i} \cdot \boldsymbol{s_j}$ where $\boldsymbol{s_i}$ is the spin operator at site $i$. The last term is the $\eta$-pseudospin term which acts between neighboring doublon and holon sites and is given by
\begin{equation}
H_{\eta}=-\frac{4t_{\text{h}}^2}{U} \sum_{\left<ij\right>} \left( \eta^x_i \eta^x_j + \eta^y_i \eta^y_j \right) -\left( \frac{4t_{\text{h}}^2}{U} - 4V \right) \sum_{\left<ij\right>} \eta^z_i  \eta^z_j , 
\label{eq_4}
\end{equation}
where $\eta^{x}_i=\delta^{A/B}(c^{\dagger}_{i\uparrow}c^{\dagger}_{i\downarrow}+h.c.)$ and $\eta^{y}_i=\delta^{A/B}(c^{\dagger}_{i\uparrow}c^{\dagger}_{i\downarrow}-h.c.)$ with $\delta^{A/B}=+1$ $(-1)$ for sublattice site $A$ $(B)$ and $\eta^{z}_i=\frac{1}{2}(n_i-1)$. When the system is completely photodoped (all the sites are converted into either doublons or holons), only $H_{\eta}$ acts and all the other terms drop out because doublons and holons have no spin. From Eq.~\eqref{eq_4} (which is the XXZ model) we obtain that if $V>2t^2_h/U$, the system goes to a charge ordered state (see Appendix B for more details).

The effective Hamiltonian for the two-orbital model captures additional physics. For $J=0$, $H_s$ is replaced by the spin-orbital term 
\begin{align}
H_{s/o}=2t_{\text{h}}^2/U \sum_{\left<ij\right>} \left( \frac{1}{2} + 2\boldsymbol{s_i} \cdot \boldsymbol{s_j} \right) \left( \frac{1}{2} + 2\boldsymbol{\tau_i} \cdot \boldsymbol{\tau_j} \right),
\end{align}
where $\boldsymbol{\tau_i}$
is the orbital pseudospin operator at site $i$. $H_{\eta}$ becomes $-4t_{\text{h}}^2/U \sum_{\left<ij\right>,\nu} \left( \eta_{i\nu}^{+}\eta_{j\nu}^{-} + \eta_{i\nu}^{-}\eta_{j\nu}^{+} \right) + V\eta^z_i \eta^z_j$ with $\nu$ an index for the six different SC orders (three spin-singlet orbital-triplet and three  spin-triplet orbital-singlet orders \cite{eta_spintrip}) and $\eta^{z}_i=\frac{1}{2}(n_i-2)$. If the two-orbital system is completely photodoped (all the sites are converted into either triplons or singlons), the effective low-energy Hamiltonian contains $H_{s/o}$ and $H_{\eta}$, because triplons and singlons have both spin and orbital degrees of freedom. If $V$ is sufficiently larger than the coefficient $4t_{\text{h}}^2/U$, $\eta^z$ becomes nonzero, which corresponds to a charge ordered state. However, now the system is also governed by the term $H_{s/o}$, which is a Kugel-Khomskii Hamiltonian with SU(4) symmetric spin-orbital structure. If the photo-doped system is cold enough (low effective temperature of the charge carriers), the charge ordered state shows Kugel-Khomskii like spin and orbital order with one sublattice occupied by triplons and the other by singlons (see sketch in Fig.~\ref{fig1}(c)).

\begin{figure}[t]
\includegraphics[width=0.43\textwidth]{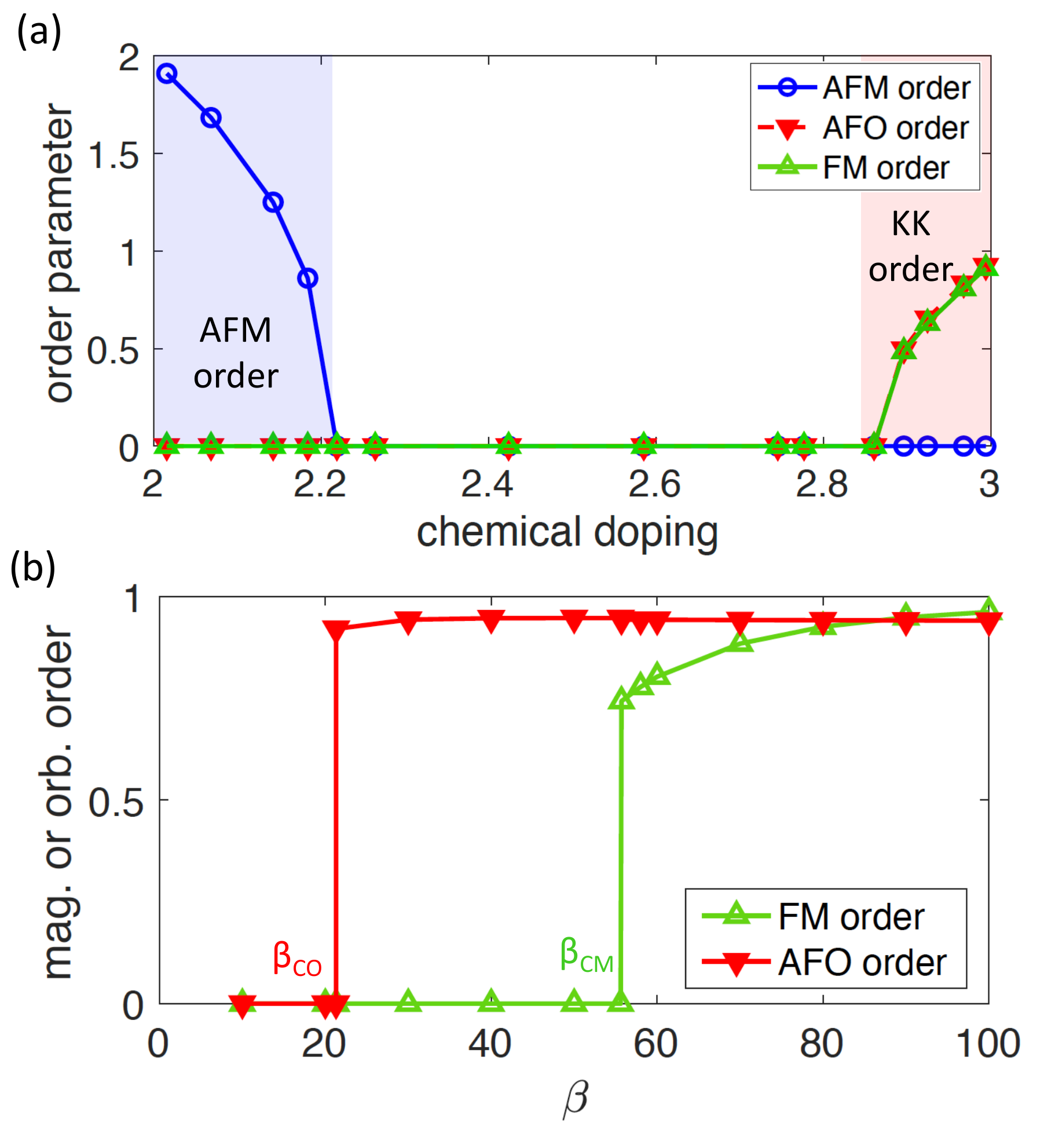}
\caption{(a) Magnetic and orbital order as a function of chemical doping for the repulsive two-orbital Hubbard model with $U=10, J=1.5$, calculated with DMFT and an NCA solver at $\beta=80$. At half-filling (chemical doping = $2$) there is AFM order (blue shaded) which melts at higher dopings, where the system enters into a paramagnetic phase. Near chemical doping $3$, Kugel-Khomskii order with FM spin and AFO order appears (red shaded region). (b) Temperature dependence of FM and AFO order at chemical doping $3$.
}
\label{fig2}
\end{figure}

\section{results}
\label{sec_results}
\subsection{Equilibrium system}

We start by studying the effect of chemical doping in the two-orbital Hubbard model. 
The ground state of the half-filled system exhibits a SU(2) antiferro order (antiferro-magnetic order for $J>0$ and antiferro-orbital order for $J<0$). In Fig.~\ref{fig2}(a), we show the magnetic order as a function of chemical doping for $U=10, J=1.5$. As we increase the chemical doping the AFM order melts and beyond a critical doping, we enter the paramagnetic phase. (For a sufficiently large $J$, there is a coexsistence  of antiferro-magnetic (AFM) and ferro-magnetic (FM) order in the intermediate region as well, consistent with the generic phase diagram in Ref.~\cite{Hoshino2016}.) As we approach $3/4$ (or equivalently $1/4$) filing, we observe Kugel-Khomskii (KK) order, which in the infinite-dimensional Bethe-lattice case corresponds to ferromagnetic spin and antiferro orbital (AFO) order (see Appendix A). The temperature dependence for the spin and orbital order is however different, as is shown in Fig.~\ref{fig2}(b). $T_c$ for the orbital order is of the order $\sim t_h^2/U$, while for the spin order $T_c \sim t_h^2 J/U^2$ (see Appendix A).

\begin{figure}[t]
\includegraphics[width=0.48\textwidth]{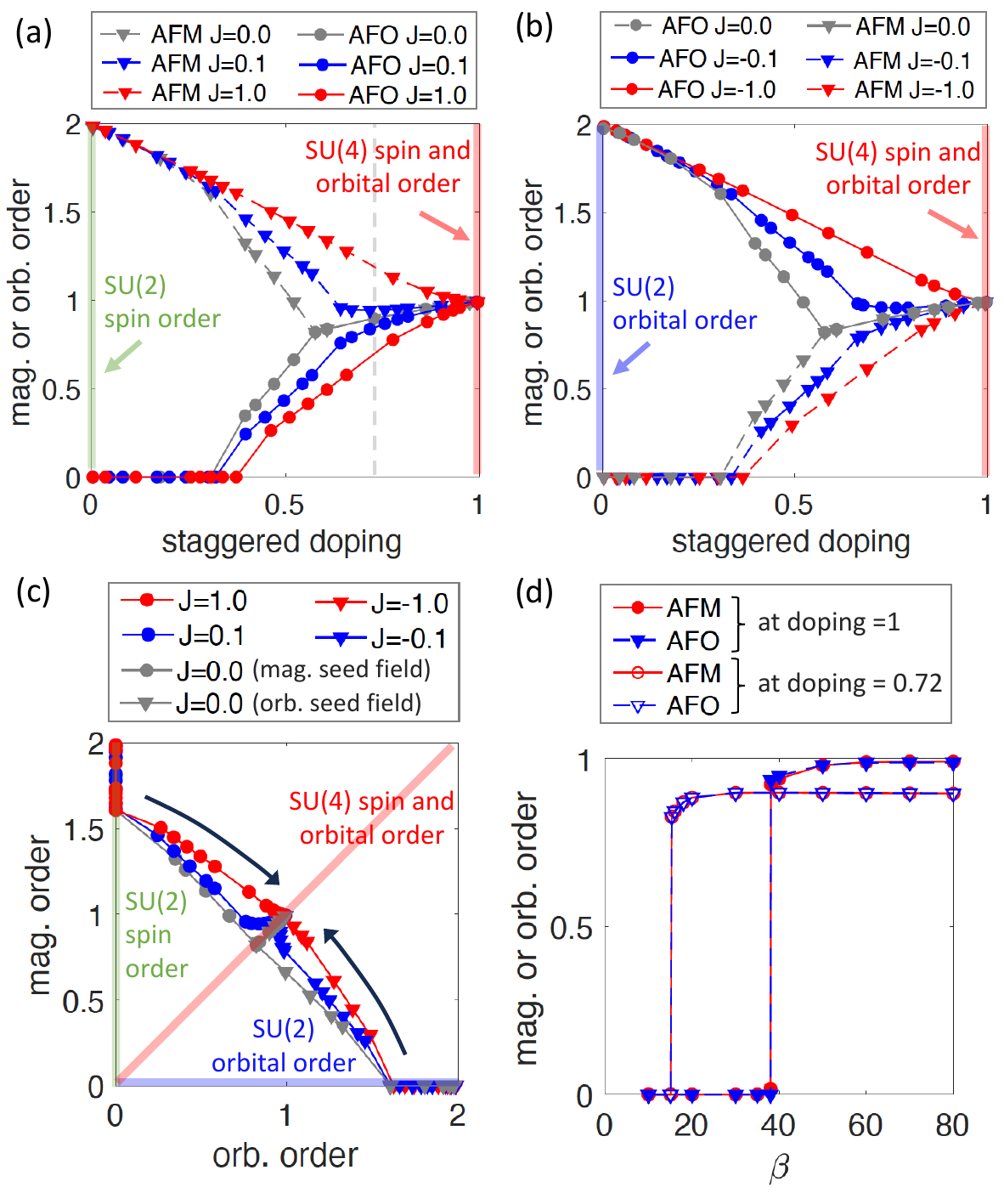}
\caption{Various orders for the two-orbital Hubbard model with staggered doping, $U=20$ and different Hund couplings $J$ at $\beta=80$. (a) Magnetic and orbital order as a function of staggered doping for $J\geq 0$. The green vertical line at staggered doping $=0$ (half-filling) represents SU(2) AFM spin order, whereas the red line at staggered doping $=1$ represents SU(4) spin and orbital order. (b) Same as (a) for $J\leq 0$. (c) Magnetic order versus orbital order plot, where the green vertical line along the $y$-axis represents SU(2) AFM spin order, the blue horizontal line along the $x$-axis represents SU(2) AFO order and the red diagonal line represents SU(4) spin and orbital order. With increasing staggered doping SU(2) order is converted to SU(4) order as indicated by the black arrows.  (d) AFM and AFO order as a function of inverse effective temperature for two different staggered dopings, 0.72 (grey dashed line in (a)) and 1 (red line in (a)).
}
\label{fig3}
\end{figure}

In this study, we are primarily interested in the emerging spin and orbital order in strongly photo-doped states, i.e., when one sublattice is triply occupied (like $3/4$ filling) and the other sublattice is singly occupied (like $1/4$ filling). To create a similar state under equilibrium conditions, we impose a staggered potential, i.e., electron dope one sublattice with a positive chemical potential shift and hole dope the other sublattice with a negative chemical potential shift.  Figure~\ref{fig3} illustrates the effect of such a staggered doping for the two-orbital repulsive Hubbard model with $U=20$ and different Hund couplings $J$. $J=0$ is a special symmetric case, where the ground state at half-filling features SU(4)-singlets in one dimension \cite{su4_1d}, SU(4)-plaquette singlets on the two-leg ladder \cite{su4_1dpl}, and a spin-orbital liquid and dimerized states in two dimensions \cite{su4_sol1,su4_sol2,su4_dimer}. In the DMFT simulations, the SU(4) symmetry can be broken down to SU(2) $\times$ U(1) by a small staggered seed field which produces a mean-field ground state. When a staggered magnetic seed field is applied, the mean-field ground state is given by AFM order and when a staggered orbital seed field is applied the mean-field ground state is given by AFO order. Figure~\ref{fig3}(a) shows the magnetic and orbital order as a function of staggered doping for $J \geq 0$. As expected, at half-filling and for $J>0$ there is AFM order and there is no orbital order. But, as we increase the staggered doping, unlike in the case of chemical doping, the AFM oder does not melt completely, but gradually decreases to a value near $1$. 

This behavior can be understood with the effective low-energy Hamiltonian described in the previous section. In the case of chemical doping, the AFM order melts because of the increasing contribution from the kinetic (hopping) term, which weakens AFM order. In the low-energy Hamiltonian the kinetic term is restricted to neighboring sites which differ in particle number by 1. However, in the staggered doping case, such a term would produce triplons in the sublattice with lower chemical potential or singlons in the sublattice with higher chemical potential, which is energetically unfavorable. Hence, the kinetic term in the Hamiltonian is suppressed and the exchange term governs the low-energy physics of the system. As a result of this, at sufficiently high staggered doping ($\sim 0.3$ for $J=0$), AFO order also starts to develop and at around 0.57 doping, both orders become equal for $J=0$ (grey curve in Fig.~\ref{fig3}(a)), which represents SU(4) spin and orbital order. For $J>0$, the onset of orbital order happens at higher doping ($\sim 0.37$ for $J=1$) and the system approaches the SU(4) spin and orbital order only at doping $=1$ (red and blue curves in Fig.~\ref{fig3}(a)). 

Interestingly, unlike in the case of chemical doping to $1/4$ or $3/4$ filling (Fig.~\ref{fig2}), for staggered doping (and equivalently for photo-doping) the Kugel-Khomskii order is antiferro in both orbital and spin. This state is favored because of a decrease in the energy cost and enhancement of the number of channels in the intermediate state of the second-order hopping processes (see Appendix A for details).

For larger values of $J$ the SU(4) symmetric order is approached more slowly. This behavior can be understood by looking at the local energy at each site. A positive Hund coupling makes the spin aligned doublon states energetically more favorable and therefore the AFM order decreases more slowly. As the doping is increased the number of doublons decreases and in the limit of staggered doping $=1$, when there are no more doublons in the system, SU(4) spin and orbital order is reached. If we flip the sign of $J$, the role of spin and orbital order reverses but qualitatively the results remain the same, which is shown in Fig.~\ref{fig3}(b). To demonstrate the interesting interplay between spin and orbital order we plot spin versus orbital order in Fig.~\ref{fig3}(c), where the $y$ and $x$ axis represents SU(2) spin and orbital order, respectively, and the diagonal direction represents SU(4) spin and orbital order. Here we clearly observe that the SU(4) order is approached from the $y$ axis for $J>0$, whereas for $J<0$ SU(4) order is approached from the $x$ axis.

Next we study the temperature dependence of the spin and orbital orders at two different dopings for $J=0$ (Fig.~\ref{fig3}(d)). Unlike in the case of chemical doping, both the spin and orbital orders now have the same $T_{c}$ of the order of $t_{h}^{2}/U$ (see Appendix A). For lower doping ($\sim 0.72$) the transition happens at a lower inverse temperature $\beta$ (higher $T_{c}$) than for doping $=1$. 
This can be understood by looking at the coefficient of the exchange term in the effective low-energy Hamiltonian \cite{SWT,eta_spintrip}, which for the triplon-singlon exchange is a factor of $1/3$ smaller compared to the doublon-doublon exchange. Hence, as the doublon number decreases, the $T_{c}$ increases. 
For $J>0$ (or $J<0$), the temperature dependence remains qualitatively the same.

\begin{figure}[t]
\includegraphics[width=0.48\textwidth]{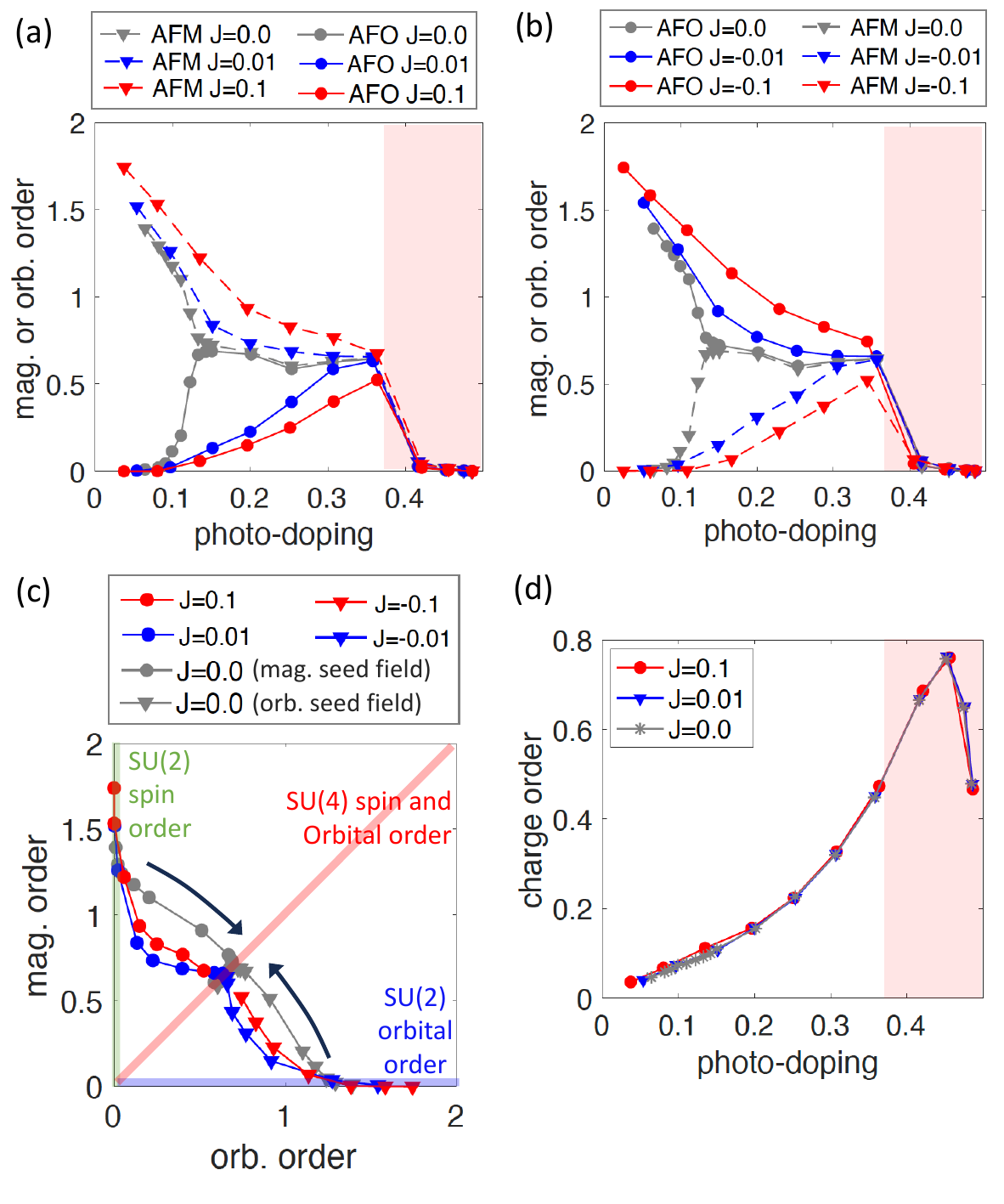}
\caption{Various orders for the photo-doped half-filled extended Hubbard model with $U=20, V=0.3$ and different Hund couplings $J$, obtained from NESS simulations. The effective temperature is kept approximately fixed around $\beta_{\text{eff}} \sim 50$. (a) Magnetic and orbital order as a function of photo-doping for $J\geq 0$. For $J=0$, a small staggered magnetic seed field is applied to break the symmtery to an AFM state in equilibrium ($0$ photo-doping). (b) Same as (a) for $J\leq 0$. For $J=0$, a small staggered orbital seed field is applied to break the symmetry to an AFO state in equilibrium.  (c) Magnetic order versus orbital order plot, where the green vertical line along the $y$-axis represents SU(2) AFM spin order, the blue horizontal line along the $x$-axis represents SU(2) AFO order and the red diagonal line represents SU(4) spin and orbital order. With increasing photo-doping SU(2) order is converted to SU(4) order as indicated by the black arrows. (d) Charge order as a function of photo-doping for different $J$. The red shaded region in (a), (b) and (d) shows the photo-doping range where we failed to reach a cold photo-doped state with the NESS approach.
}
\label{fig4}
\end{figure}

\subsection{Nonequilibrium system}

\begin{figure*}[t]
\includegraphics[width=0.9\textwidth]{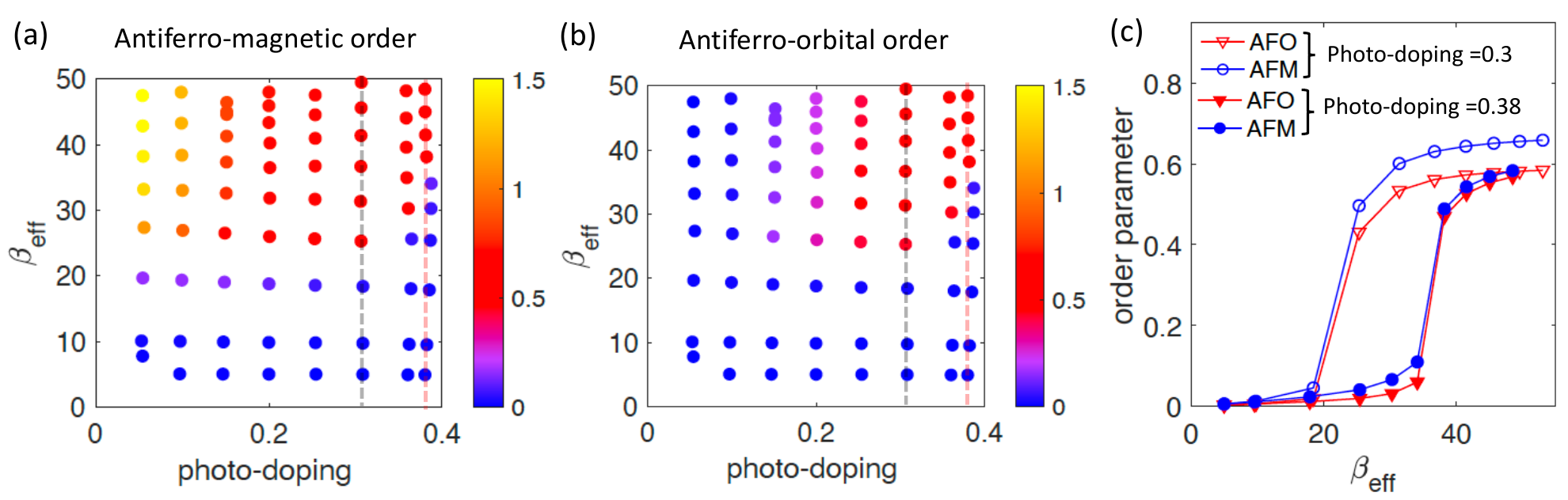}
\caption{Temperature dependence of AFM and AFO order for $U=20, V=0.3$ and $J=0.01$. (a) AFM order and (b) AFO order as a function of inverse effective temperature $\beta_\text{eff}$ and photo-doping (color map). The Kugel-Khomskii order (red color) is reached at photo-doping $\gtrsim 0.25$ and sufficiently high $\beta_\text{eff}$ (low $T_\text{eff}$). (c) AFM and AFO order as a function of $\beta_\text{eff}$ for two different photo-dopings as indicated by the grey (photo-doping $=0.3$) and red (photo-doping $=0.38$) dashed lines (a) and (b).
CO is stable up to the highest effective temperatures considered.
}
\label{fig5}
\end{figure*}

The effects of staggered doping are qualitatively reproduced in the photo-doped half-filled extended Hubbard model with nearest-neighbor repulsive interaction $V$. In Fig.~\ref{fig4}, we show the results of the NESS calculations for the photo-doped extended Hubbard model with $U=20$ and $V=0.3$. As mentioned earlier, a small nearest-neighbor interaction $V$ is enough to drive the system to a charged ordered state. We quantify the amount of photo-doping in this charge ordered state by taking the average of the triplon densities (or equivalently the singlon densities) in sublattices $A$ and $B$. In this notation a photo-doping of $0.5$ indicates complete population inversion, i.e., all the sites are converted into triplons and singlon. In Fig.~\ref{fig4}(d) we show the charge order ($|c_{A}|$ or $|c_{B}|$) as a function of photo-doping. As the photo-doping increases, more charge excitations (triplons and singlons) are produced and they become charge ordered because of the repulsive $V$. In particular, the charge order increases with photo-doping almost independently of the Hund coupling $J$. In Fig.~\ref{fig4}(a) and (b) we show how spin and orbital order emerges with photo-doping for $J\geq 0$ and $J\leq 0$. For $J=0$ we apply a small staggered magnetic seed field in Fig.~\ref{fig4}(a) and a staggered orbital seed field in Fig.~\ref{fig4}(b) to break the symmetry. As the photo-doping is increased for $J=0$,  with a magnetic seed field, we observe that the AFM order does not melt completely and an AFO order starts to develop at photo-doping $\sim 0.09$ and becomes almost equal to the AFM order at photo-doping $\sim 0.15$, which indicates an approximate SU(4) spin and orbital order. Such ordering can be explained by a similar mechanism as for the staggered doping case described in the previous section. The kinetic term is heavily suppressed by the charge ordering, which implies that the photo-doped system is governed by the exchange term of the effective Hamiltonian. 

For $J>0$, the AFM order decreases more slowly compared to $J=0$. This is because the positive Hund coupling favors spin-aligned doublons, which stabilize the AFM order. As the photo-doping is increased, the number of doublons decreases, which weakens this effect, so that the system approaches SU(4) spin and orbital order at high photo-doping. For $J<0$ the roles of the spin and orbital order are reversed, as expected \cite{Steiner2016}, which is illustrated in Fig.~\ref{fig4}(b). For larger photo-doping or for higher values of $J$, it becomes challenging to converge the self-consistent solution at an effectively low temperature, and above photo-doping $\sim 0.4$ (indicated by the red shaded region), we are unable to obtain cold photo-doped states. For this reason the spin and orbital orders disappear above photo-doping $>0.4$. In Fig.~\ref{fig4}(c), we show the magnetic versus orbital order plot for different $J$ and photo-doping. Similarly to the staggered doping case, we observe that the SU(2) orders from the $y$ and $x$ axes approach the SU(4) order on the diagonal for $J>0$ and $J<0$, respectively.

 Finally, we study the temperature dependence of the spin and orbital order in the photo-doped system  with $J=0.01$ (Fig.~\ref{fig5}). The color map shows the AFM order in Fig.~\ref{fig5}(a) and the AFO order in Fig.~\ref{fig5}(b) as a function of photo-doping and effective inverse temperature $\beta_\text{eff}$ (the results remain qualitatively the same for different $J$). We observe that both orders start appearing for $\beta_\text{eff}>20$ (below an effective temperature $\sim t_{h}^{2}/U$). For higher photo-doping and lower effective temperature, SU(4) spin and orbital order starts to develop (red color). In Fig.~\ref{fig5}(c), we plot the AFM and AFO orders as a function of effective temperature for the two different photo-dopings indicated by the grey and red dashed lines in Fig.~\ref{fig5}(a) and (b). As in the staggered-doping case, $T_{c}$ is reduced for higher photo-doping, which shows that the same mechanism holds in the photo-doping and staggered doping case.

\section{Conclusions}
\label{sec_conclusion}
We demonstrated that intertwined spin and orbital order of the Kugel-Khomskii type can be photo-induced in the Mott insulating two-orbital extended Hubbard model. Both the equilibrium DMFT calculations with staggered doping and the NESS simulations of the half-filled photo-doped system produce AFM and AFO ordering as we approach the high doping limit. This is in contrast to the combined FM and AFO order found in the uniformly doped equilibrium system near 3/4 or 1/4 filling. This qualitative difference could be explained by analyzing the second-order hopping processes and the energy cost for the intermediate states involved in the exchange interactions for both the equilibrium and non-equilibrium cases. Another peculiarity of the photo-doped system is that the Kugel-Khomskii type order emerges out of a nonthermal charge ordered state, which is stabilized by the repulsive inter-site interaction. 

We also showed that the two-orbital $U$-$V$ Hubbard model provides an interesting example where we can tune the SU(2) spin order (or orbital order) at half-filling to an SU(4) spin and orbital order using photo-doping as a control knob. The coexistence of charge order along with the Kugel-Khomskii order demonstrates the possibility of realizing rich intertwined hidden orders in multi-orbital systems. Higly tunable hidden orders, especially magnetic orders, could provide a basis for future ultrafast devices. 

Entropy cooling techniques \cite{entropy_cooling} with optimized chirped pulses \cite{chirped_pulse} can be used to overcome the challenge of realizing effectively cold photo-doped systems. Cold atoms in optical lattices may provide an experimental platform for realizing highly photo-doped Mott states in various models, since a long life-time of photo-carriers has been demonstrated in such systems systems \cite{Sensarma2010,Strohmaier2010}. Recently proposed cold atom techniques for implementing multi-orbital physics \cite{richaud2022} should enable the realization and detection of the hidden orders proposed in this study.


\section{Acknowledgements} 
This work has been supported by SNSF Grant No. 200021-196966. The NESS simulations are based on a code originally developed by J.~Li and M.~Eckstein. The calculations were run on the beo06 cluster at the University of Fribourg.

\appendix
\section{Kugel-Khomskii model}
Here we describe the details of the Kugel-Khomskii model for the equilibrium system with $1/4$ or $3/4$ filling and for the photo-doped half-filled system. Kugel-Khomskii order can be described by introducing spin $\boldsymbol{s}$ and orbital pseudo-spin $\boldsymbol{\tau}$ operators \cite{Kugel_Khomskii}. For $J=0$ (no Hund coupling), the Kugel-Khomskii Hamiltonian is given by
\begin{eqnarray}
H_{KK}=\frac{t_{\text{h}}^2}{U} \sum_{\left<ij\right>} \left( \frac{1}{2} + 2\boldsymbol{s_i} \cdot \boldsymbol{s_j} \right) \left( \frac{1}{2} + 2\boldsymbol{\tau_i} \cdot \boldsymbol{\tau_j} \right),
\end{eqnarray}
which has a SU(4) `spin' structure (note that in our notation $\boldsymbol{s}$ always indicates a  SU(2) spin). In this symmetric case, the ground state exhibits a complicated ordering pattern as described in Sec.~\ref{sec_results}. A nonzero Hund coupling $J$ breaks the SU(4) symmetry into a SU(2)$\times$SU(2) symmetry and the corresponding Hamiltonian is given by
\begin{eqnarray}
  H_{KK}&=&\frac{t_{\text{h}}^2}{U} \sum_{\left<ij\right>} \Bigg[ \Big( \frac{1}{2} + 2\boldsymbol{s_i} \cdot \boldsymbol{s_j} \Big) \Big( \frac{1}{2} + 2\boldsymbol{\tau_i} \cdot \boldsymbol{\tau_j} \Big)  \nonumber \\
 &&\hspace{1 cm} + \frac{J/U}{1-(J/U)^2} \Big[ 2( \boldsymbol{\tau_i} \cdot \boldsymbol{\tau_j} -  \tau_i^z \tau_j^z ) \nonumber \\
 &&\hspace{1 cm} -\Big( \frac{1}{2} + 2\boldsymbol{s_i} \cdot \boldsymbol{s_j} \Big) \Big( \frac{1}{2} - 2\tau_i^z \tau_j^z \Big)\Big]\Bigg].\hspace{3mm}
 \label{eq_A2}
\end{eqnarray}
In our case, the $z$ component of the orbital pseudo-spin $\boldsymbol{\tau}$ represents the orbital occupation ($\tau^z=1/2$ indicates that orbital $1$ is occupied and $\tau^z=-1/2$ indicates that orbital $2$ is occupied). The Hamiltonian in~Eq.~(\ref{eq_A2}) suggests a ferromagnetic spin state and and antiferro orbital pseudo-spin state at zero temperature. As temperature is increased, the magnetic order vanishes at $T\sim t_{\text{h}}^2J/U^2$ and orbital order at $T\sim t_{\text{h}}^2/U$,  as described in Refs.~\cite{Kugel_Khomskii} and \cite{Kugel_Khomskii_2}. 

\begin{figure}[t]
\includegraphics[width=0.48\textwidth]{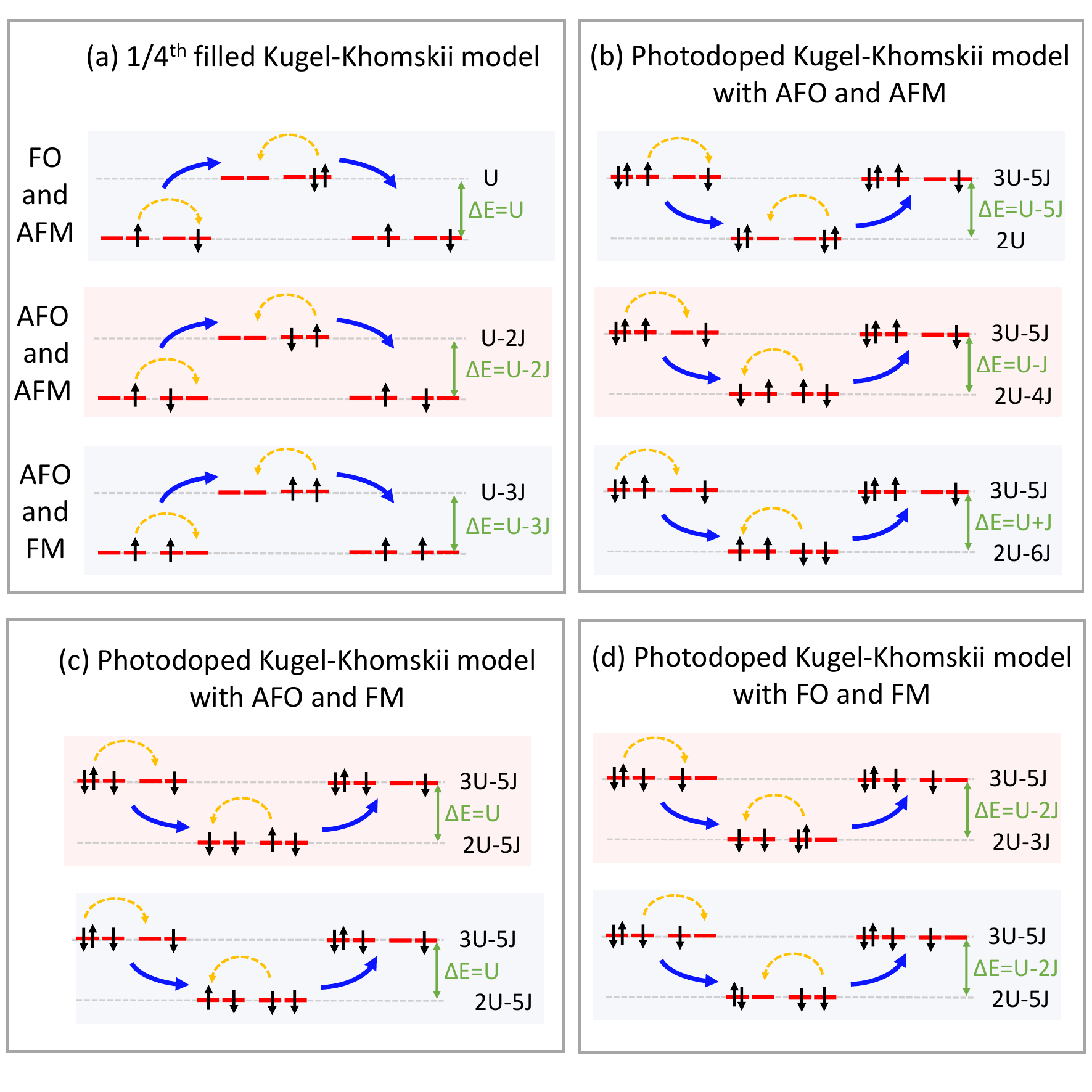}
\caption{Various second order hopping processes contributing to different orbital and magnetic orders in equilibrium and in photodoped systems. (a) Second order hopping processes in the $1/4$ filled Kugel-Khomskii model. $\Delta E$ indicates the energy difference between the intermediate and the initial (or equivalently final) state. (b-d) Second order hopping processes in the photodoped Kugel-Khomskii model. Only AFO and AFM order has three intermediate hopping channels, as shown in (b). Other orders have only two possible hopping channels and FO and AFM order would be similar to FO and FM order shown in (d).
}
\label{fig_A1}
\end{figure}

This ordering can also be qualitatively understood by considering the intermediate states in the second order hopping processes for the exchange interactions. In ~\figref{fig_A1}(a), we show all the possible orbital and magnetic orders for the Kugel-Khomskii model representing the $1/4$ filled two-orbital Mott system. In this case FO and FM order is not possible because there are no corresponding intermediate states.  Among the possible orders, AFO and FM order have the intermediate state which costs the minimum energy $U-3J$. Therefore, this order will emerge as the Kugel-Khomskii order in the chemically doped equilibrium system, as shown in ~\figref{fig2}(a). 

Both the AFO and AFM exchange terms have an exchange coefficient of the order of $\sim t_{\text{h}}^2/U$ whereas the FM exchange coefficient is of the order $\sim t_{\text{h}}^2J/U^2$ (third term in ~\eqref{eq_A2}). Since the AFO order is more robust, we can plug in $\boldsymbol{\tau_i} \cdot \boldsymbol{\tau_j}=\tau_i^z \tau_j^z=-1/4$ for AFO order in ~\eqref{eq_A2}. Then, the effective spin Hamiltonian takes the form (neglecting the constant terms)
\begin{eqnarray}
H_{KK,s}=-\frac{2t_{\text{h}}^2}{U}\cdot \frac{J/U}{1-(J/U)^2} \sum_{\left<ij\right>} \boldsymbol{s_i} \cdot \boldsymbol{s_j},
 \label{eq_A3}
\end{eqnarray}
which gives us the estimate for the FM transition temperature as $\sim t_{\text{h}}^2J/U^2$.

A similar analysis of the intermediate states reveals that a different order is favored in the photodoped system, as shown in ~\figref{fig_A1}(b-d). For $J=0$ all the intermediate hopping processes are degenerate, but the AFO and AFM order has the maximum number of intermediate hopping channels. So, in the absence of Hund coupling, such an order should be favored. For nonzero $J$, the degeneracy in the intermediate processes is lifted, but the minimum energy cost $U-5J$ again favors the AFO and AFM order. Since both AFO and AFM order have the same exchange coefficient ($\sim t_{\text{h}}^2/U$), in this case, both orders show a similar temperature dependence and the same transition temperature.

\begin{figure*}[t]
\includegraphics[width=0.95\textwidth]{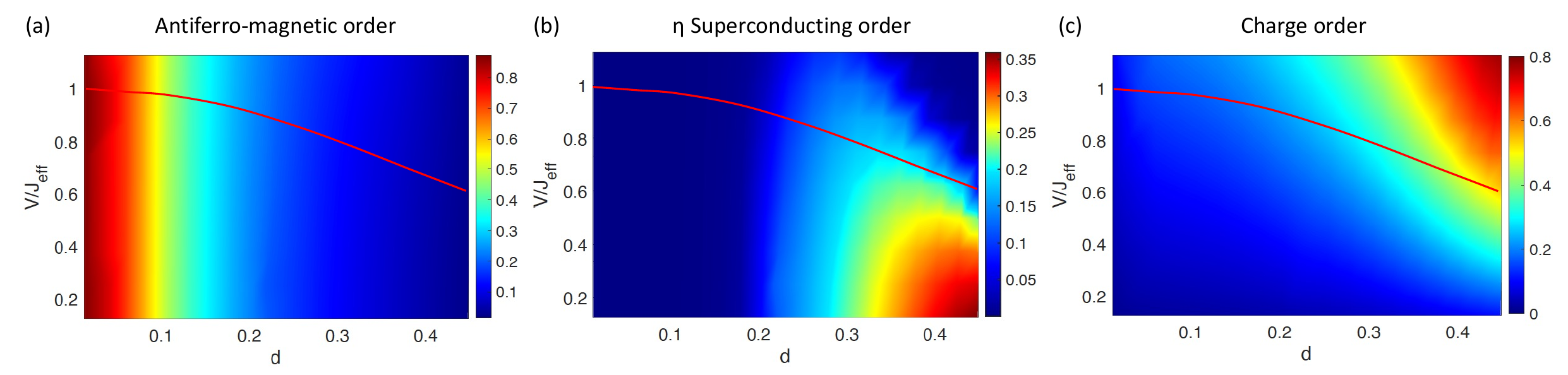}
\caption{Different order parameters for the one-orbital Hubbard model with $U=10$ at $\beta_{\text{eff}} \sim 20$. (a) AFM order, (b) $\eta$-SC order and (c) charge order as a function of photo-doping and nearest neighbor repulsion $V$ (color map). The red line indicates the phase boundary between the $\eta$-SC state and the charge order state (at nonzero photo-doping) obtained for the one-dimensional model at effective temperature zero with iTEBD \cite{yuta_etasc2}.
}
\label{fig_B1}
\end{figure*}

\section{One-orbital extended Hubbard model}
In this section, we briefly discuss our NESS DMFT calculations for the single-orbital $U$-$V$ Hubbard model. To simulate a photo-doped system, we consider the model with $U=10$ and place two cold Fermion baths at $\omega_{b}=\pm U/2$ with $W_{b}=2.5$ and coupling strength $\Gamma=0.06$. The nearest-neighbor repulsion term $V$ is treated at the mean-field level, as discussed in Sec.~\ref{sec_model}. The amount of photo-doping is measured by the doublon density $d=\langle n_{\uparrow}n_{\downarrow}\rangle$. As discussed in Sec.~\ref{sec_eff_lowen_model}, the effective low-energy Hamiltonian in the limit of complete photodoping is given by Eq.~\eqref{eq_4}, which can be identified as the XXZ model, where the $x$ and $y$ components correspond to staggered superconducting order ($\eta$-SC) and the $z$ component to charge order. This model has been studied with the help of DMRG techniques \cite{yuta_etasc1,yuta_etasc2}. The ground state of the single-orbital Hubbard model exhibits AFM order. If the system is photo-doped, the AFM order disappears and for $V=0$ the system goes to an $\eta$-SC state at zero effective temperature. In the presence of a nonzero $V$, there is a competition between $\eta$-SC and charge order. The phase diagram of this model in one dimension as a function of photo-doping and nearest-neighbor repulsion $V$ has been mapped out using the time-evolving block decimation (iTEBD) and exact diagonalization (ED) \cite{yuta_etasc1,yuta_etasc2}. For small photo-doping, the critical $V_{\text{c}}$ is $J_{\text{eff}}$ ($J_{\text{eff}}=4t_{\text{h}}^2/U$). With increasing photo-doping,  $V_{\text{c}}$ gradually decreases and in the complete photo-doping limit $V_{\text{c}}=J_{\text{eff}}/2$.

We performed NESS simulation for the same model on the infinitely connected Bethe lattice at nonzero temperature. Interestingly, our DMFT phase diagram agrees rather well with the previous iTEBD results for the 1D system, which indicates a common underlying ordering mechanism in both lattices and also validates our mean-field approach. Figure~\ref{fig_B1} shows the different order parameters using a colormap in the space spanned by photo-doping (doublon density) and $V$. As expected, at zero photodoping, there is an AFM phase, which melts with increasing photo-doping due to the kinetic term and we enter into a paramagnetic phase as shown in ~\figref{fig_B1}(a). 
When we increase the photo-doping further, above $d\sim 0.3$ both charge order (for higher values of $V$) and  $\eta$-SC order (for lower values of $V$) appear, as shown in ~\figref{fig_B1}(b) and (c). The red line shows the $V_{\text{c}}$ obtained for the one-dimensional system with iTEBD \cite{yuta_etasc2}, which is roughly consistent with our phase diagrams in ~\figref{fig_B1}. Our calculations are for nonzero effective temperature ($\beta_\text{eff}\approx 20$), which explains the melting of the orders at intermediate doping due to the kinetic term. Such a melting is absent in the previous iTEBD studies at zero effective temperature, where there is a direct phase transitions between the three states (AFM, $\eta$-SC and charge order) without any intermediate disordered phase.

\end{document}